\documentstyle[twocolumn,epsf,aps]{revtex}
\draft
\begin{document}

\title{Vortex dynamics in a three-state model \\
        under cyclic dominance}
\author{Gy\"orgy Szab\'o$^{1,2}$, M. A. Santos$^2$, and 
J. F. F. Mendes$^2$}

\address
{$^1$ Research Institute for Technical Physics and Materials Science \\
POB 49, H-1525 Budapest, Hungary}

\address
{$^2$ Departamento de F\'{\i}sica and Centro de F\'\i sica do Porto \\
Faculdade de Ci\^encias, Universidade do Porto \\
Rua do Campo Alegre 687, 4150 Porto, Portugal}

\address{\em \today}

\address{
\centering{
\medskip \em
\begin{minipage}{15.4cm}
{}~~~The evolution of domain structure is investigated in a two-dimensional
voter model with three states under cyclic dominance. The study focus on
 the dynamics of vortices, defined by the points where the three states
(domains) meet. We can distinguish vortices and antivortices which walk
randomly and annihilate each other. The domain wall motion can
create vortex-antivortex pairs at a rate which is increased by the
spiral formation due to cyclic dominance. This mechanism is contrasted
with a branching annihilating random walk (BARW) in a particle-antiparticle
system with density-dependent pair creation rate. Numerical estimates
for the critical  indices of the vortex density ($\beta = 0.29(4)$) and
of its fluctuation ($\gamma=0.34(6)$) improve an earlier Monte Carlo
study [Tainaka and Y. Itoh, Europhys.\ Lett.\ {\bf 15}, 399 (1991)]
of the three-state cyclic model in two dimensions.
\end{minipage}
}}
\maketitle

\narrowtext

The self-organizing domain structures in the cyclic variants of the
Lotka-Volterra model \cite{Lotka,Volterra} have been extensively investigated
because similar spatio-temporal oscillations can appear in chemical reactions
as well as in more complex ecological processes. These
phenomena can be well studied within the formalism of voter models 
\cite{Ligget} we will follow henceforth. Different versions of the
three-state voter models under cyclic dominance on a square lattice were
introduced by Itoh and Tainaka \cite{Itoh,TI,Tainaka}. Bramson and Griffeath
\cite{BG} studied the flux and fixation in cyclic particle systems.
The pattern formation in cyclic cellular automata was investigated by
Fisch \cite{Fisch}. Using a pair approximation, Frachebourg and 
Krapivsky \cite{FK} have shown that fixation occurs in a cyclic 
Lotka-Volterra model if the system is started from a random
initial state and the number of states
exceeds a critical value dependent on the dimension. According to
this result, the system tends toward a self-organizing, inhomogeneous
state if the number of states is less than 14 on a square lattice.
Unfortunately, the theoretical understanding of the mechanism maintaining
the inhomogeneous state is still incomplete.

In this work we consider the features of a self-organizing domain structure
in a two-dimensional system using the concept of vortices defined for
three-state models. Instead of studying the average sizes of the domains,
Tainaka and Itoh \cite{TI} have determined the average density of vortices 
($c=\langle N_v \rangle /L^2$ where $N_v$ denotes the number of vortices in
a system with $L \times L $ lattice points). The vortices are points in these
domain structures where the three different states ($A$, $B$ and $C$) and
the three types of domain walls meet. Evidently, the value of $c^{-1}$
represents roughly the average domain area (size). The investigation of
vortex density was strongly motivated by the fact that its determination
is much easier than the evaluation of the average domain size.

\begin{figure}
\centerline{\epsfxsize=7cm
            \epsfbox{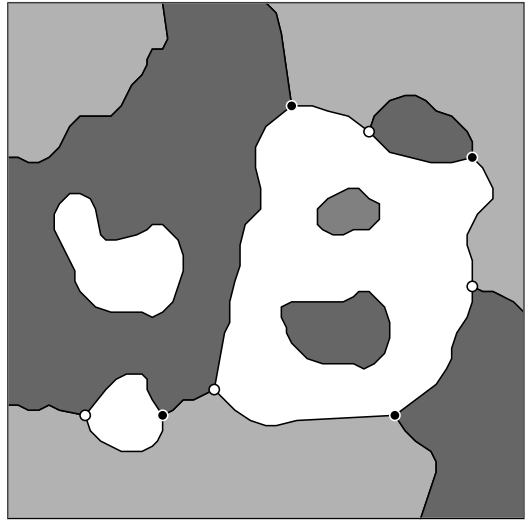}
            \vspace*{1mm}   }
\caption{A typical domain structure for a three-state (color) model.
The black bullets with white border and the white bullets with
black border represent the vortices and antivortices.}
\label{fig:map3c}
\end{figure}

Figure 1 shows a (three-color) domain structure on the macroscopic scale.
It is easy to recognize that two types of vortices may be distinguished
as indicated by black and white bullets in the figure. We will call them
vortex and antivortex depending on whether we find $ABC$ or $ACB$ order
when going clockwise around the center. A simple rule may be deduced at
first glance, namely the number of vortices is even around a closed domain.
More precisely, the vortices and antivortices are alternately located
along the closed boundary of a domain. This feature has serious consequences
when the motion and collision of vortices is considered.

\begin{figure}
\centerline{\epsfxsize=6cm
            \epsfbox{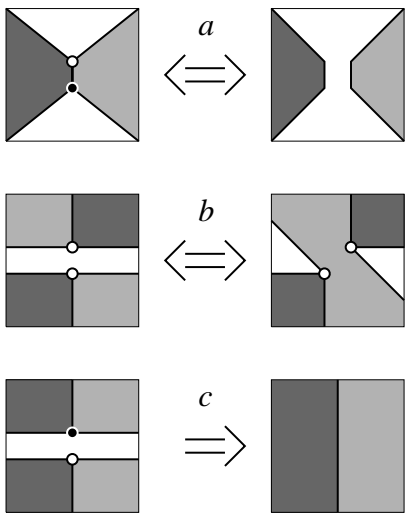}
            \vspace*{1mm}   }
\caption{Shematic plots for characterizing the elements of vortex dynamics
on a three-state map. The upper process ($a$) represents the annihilation
of a vortex-antivortex pair walking along their common boundary line
as well as the reverse phenomenon corresponding to a spontaneous pair
creation. ($b$) The ``collision'' of two vortices or antivortices can modify
the type of domain separating them. ($c$) A vortex-antivortex pair
can annihilate each other in a different way when crossing through
the separating domain.}
\label{fig:vordyn}
\end{figure}

During the time evolution of a three-color domain structure the vortices
move together with the boundaries. In these processes the vortices can
collide and annihilate each other. Figure \ref{fig:vordyn} illustrates
the typical elementary events (after the transient processes) whose
combinations describe all the possible phenomena related to the creation,
annihilation and collisions of the vortices. In the present approach the
topological situations where four (or more) domain boundaries meet at a
given point are considered as instantaneous events of collisions, pair
annihilations or creations. Evidently, the illustrated elementary events
modify the connectivity among the vortices. This connectivity, however,
can be modified by either fusion or fission of domains without any change
in their constellation.

A numeric analysis of the vortex density was performed by Tainaka and Itoh
\cite{TI} in a two-dimensional voter model where the voters, located on a
square lattice, should choose among three
states (opinions): $A$, $B$ and $C$. The system evolution is governed by a
simple algorithm: a randomly chosen voter or one of its nearest
neighbors can modify their opinion if those are different. If the chosen
voter and its neighbor are in the states $A$ and $B$, then the first
voter adopts opinion $B$ with probability $P$, otherwise its neighbor
changes its state to $A$. This adoption rule is repeated cyclically for the
$B$-$C$ and $C$-$A$ cases.
Similar cyclic dominance characterizes the ``paper, scissors, stone''
games \cite{games}.

The direction of dominance can be reversed by replacing $1-P$ for $P$,
therefore the analyses are restricted to $P > 1/2$. For $P=1/2$
the present model is equivalent to a traditional voter
model \cite{Ligget} which exhibits a (three-color) domain coarsening
phenomenon if initially the voter states are random. In other words,
the finite system evolves into one of the three homogeneous states
while the vortex density goes to zero.

For $P>1/2$, however, a self-organizing domain structure is maintained,
and the vortex density tends to a stationary value, $N_v$
dependent on $P$. More precisely, Tainaka and Itoh have found a power
law behavior, namely,
\begin{equation}
c \sim (P-P_c)^{\beta} \
\label{eq:beta}
\end{equation}
where $P_c=1/2$ and $\beta \sim 0.40$ \cite{TI}.

We have repeated these simulations using larger system size ($400 \times 400$)
and longer sampling times. In the vicinity of $P_c$
($ P-P_c < 0.01$) the thermalization is chosen to be longer than
250,000 MCS (Monte Carlo Steps per particle). In the stationary state
we have determined all the four-point configuration probabilities 
$Q(n_1,n_2,n_3,n_4)$ [$n_i=A, B$ or $C$; ($i=1,2,3,4$)] of a
$2 \times 2$ cluster. In our notation
the $(A,A,B,C)$ configuration refers to a vortex, the $(A,A,C,B)$ to an
antivortex, and the $(A,B,C,A)$ configuration can be interpreted as a
vortex-antivortex pair (before their annihilation or after their birth).
Monitoring the vortex density we were able to determine its average value
and its fluctuation [$ \chi = L^2 \langle (N_v/L^2 - c)^2 \rangle $] 
in the stationary states as a function of $p \equiv P-P_c$.

\begin{figure}
\centerline{\epsfxsize=8cm
            \epsfbox{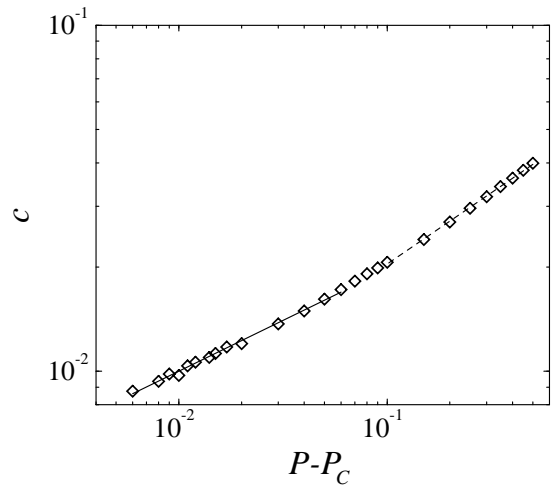}
            \vspace*{1mm}   }
\caption{Log-log plot of vortex density {\it vs} $P$ in the three-candidate
voter model under cyclic dominance. The open diamonds represent MC data,
the solid and dashed lines (resp. slopes  0.29 and 0.41) indicate the
fitted power laws.}
\label{fig:v3c}
\end{figure}

\begin{figure}
\centerline{\epsfxsize=8cm
            \epsfbox{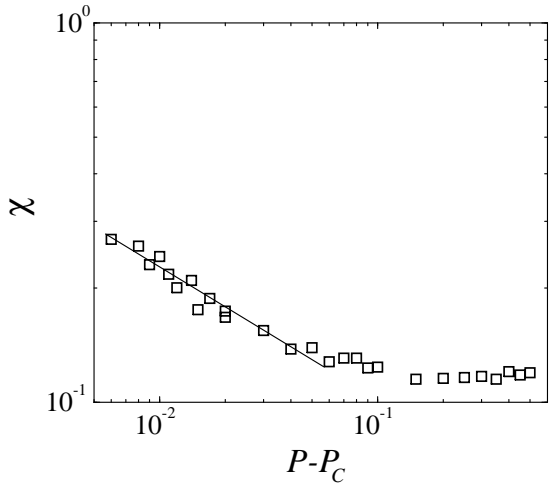}
            \vspace*{1mm}   }
\caption{Log-log plot of density fluctuation {\it vs} $P-P_c$ in the 
three-candidate voter model under cyclic dominance. MC results are
represented by open squares, the fitted power law function is represented
by a solid line for small $P-P_c$ values.}
\label{fig:v3cf}
\end{figure}

The results of our simulations are summarized in two log-log plots 
(see  Figs.~\ref{fig:v3c} and \ref{fig:v3cf}). Figure \ref{fig:v3c}
demonstrates that we have reproduced Tainaka and Itoh's data
\cite{TI} for $p > 0.1$. In this region the fitted power law 
(dashed line) is characterized by an exponent $\beta = 0.41(3)$ in
agreement with Tainaka and Itoh. For smaller $p$ values, however, we
have found a different exponent $\beta = 0.29(4)$ (solid line).
Similar crossover behavior can be observed 
for the fluctuation as shown in Fig. \ref{fig:v3cf}. The fluctuation
remains approximately constant for $p > 0.1$, whereas it follows a power
law ($\chi \sim (P-P_c)^{- \gamma}$ with $\gamma = 0.34(6)$)
for smaller $p$ values.

The above mentioned configuration probabilities satisfy the reflection,
rotation of $90^{\circ}$ and cyclic symmetries. Consequently, all these
quantities are describable, within a cluster approximation by introducing
6 independent parameters
which can be evaluated by solving a set of equations of motion for
the four-point configuration probabilities. Previously, this method
has been proved to be a very efficient tool for the investigations of
stochastic cellular automata \cite{CA}, evolutionary games \cite{PD},
different lattice versions of the Lotka-Volterra models (at the
level of pair approximation) \cite{Tainaka,FK}, and for a two-dimensional
driven lattice gas maintaining a self-organizing domain structure too
\cite{sods}. In the present case, the calculations show
a very weak $P$-dependence of the configuration probabilities including the
vortex constellations mentioned above --- in contrast with the MC
simulations. At the level of the pair approximation 
the configuration probabilities are independent of $P$ \cite{Tainaka},
while the mean-field (1-point) approximation predicts homogeneous
oscillatory behavior. At the five-point level, on the other hand,
the preliminary calculations (integrating numerically the equations of
motion) also indicate a very weak $P$-dependence. This puzzling failure
of the dynamical cluster technique inspired us to search for a mechanism
observable at macroscopic (or mesoscopic) scale. 

\begin{figure}
\centerline{\epsfxsize=8cm
            \epsfbox{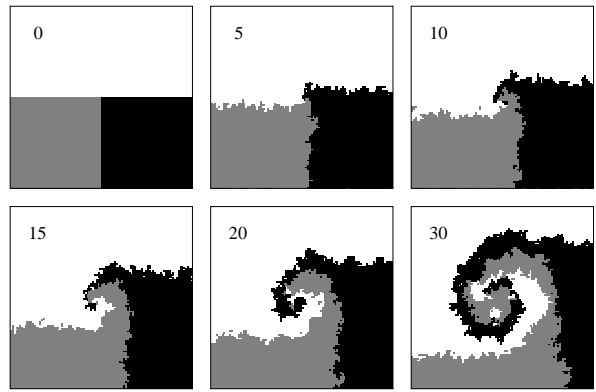}
            \vspace*{1mm}   }
\caption{Time evolution of a vortex initially having straight border lines
for $P=1$. The figures at the upper-left corners indicate the time measured
in MCS units.}
\label{fig:cfgs}
\end{figure}

For $P=1/2$, the domain coarsening is accompanied by the annihilation of
pairs, and this process is not prevented by the weak spontaneous pair
creation. Under cyclic dominance ($P>1/2$), however, we have detected
the appearence of a pair creation mechanism which is able to compensate
for the previous annihilation process, yielding a finite vortex density.
We have displayed in Fig. \ref{fig:cfgs} the evolution of a single vortex
whose geometry allows us to recognize the essential processes.
Here, instead of the usual periodic boundary
conditions, we have assumed that a voter residing on the perifery and its
''hypothetic'' outer neighbor have always the same opinion. Along the
boundaries one can observe a cyclic invasion whose average velocity component
perpendicular to the border line is proportional to $P-P_c$. This motion
of the boundaries yields a spiral formation around the center.
The average time evolution is decorated by noise as shown in a series
of snapshots in Fig. \ref{fig:cfgs}. Due to the randomness the neighboring
boundaries can contact and create vortex-antivortex pairs inside the
spiral because it consists of narrow "arms" which is a favoured situation
for the pair creation. The created pairs can be considered as the offspings
of the original vortex. Most of the pairs are annihilated within a short time
but, sooner or later, a vortex-antivortex pair will eventually drift apart
from each other. The corresponding vortices will then expand and become
capable to create further offsprings via the same spiral formation process.
The self-organizing domain structure can thus be maintained by this mechanism.

Within the framework of the vortex language, the evolution of domain structure
can be described by the two-dimensional motion of vortices allowing the
annihilation and creation of pairs. Assuming that the motion of vortices
is dominantly controlled by noise, the present problem can be considered
as a so-called branching annihilating random walk (BARW) with two types of
particles created and annihilated in pairs. The branching process is
evidently controlled by the value of $P-P_c$, though the mathematical
relation between the branching rate and $P-P_c$ is not yet clarified.
Furthermore, the branching rate is affected by the nearest neighbor
distances because it limits the spiral formation.

The traditional BARWs have been intensively studied in the last years
(for recent reviews see the work by Cardy and T\"auber \cite{CT} and
Marro and Dickman \cite{Dickman}) because they undergo a critical
transition when varying the rate of branching. The corresponding critical
behavior belongs to the DP (directed percolation) universality class
\cite{dp} involving the Reggeon field theory \cite{reggeon}, the surface
reaction \cite{surfreact} and Schl\"ogl models \cite{Schlogl}, and the
extinction phenomena observed in spatial evolutionary games \cite{PD}.
According to the "DP conjecture" \cite{dpconj} most of the one-component
model with a single absorbing state belong to the DP universality class.
Exceptions can appear when additional symmetries \cite{haye} or conservation
laws are introduced. Well known examples are those models in which the
parity of particles is conserved during the elementary processes \cite{TT,PC}.
The introduction of two (or more) types of particles has enlarged the
number of possible universality classes \cite{CT,HT}.

The field theoretical results \cite{CT,HT,LC} indicate that mean-field
approaches can give satisfactory description of the two-dimensional BARW
models with two types of particles. This observation motivated us to try
to describe the branching annihilating random walks of vortices and
antivortices through a mean-field equation with a density-dependent
branching rate, namely
\begin{equation}
{\partial \over \partial t} c = - c^2 + \lambda (c) c
\label{eq:mf}
\end{equation}
where $c$ denotes the concentration of vortices and antivortices.
The first term describes the annihilation process whose prefactor is
eliminated by choosing a suitable time scale. For simplicity we
suppose that the density dependence of the branching rate follows 
a power law with exponent $\nu$
\begin{equation}
\lambda (c)= p c^{-\nu}
\label{eq:lambda}
\end{equation}
where $p=A (P-P_c)$. Notice that this branching rate diverges in the
limit $c \to 0$ if $\nu > 0$. In the stationary state one can easily
determine the concentration as a function of $p$:
\begin{equation}
c \sim p^{1 \over 1 + \nu } \ .
\label{eq:nu}
\end{equation}
Comparing this formula with the above MC results one can conclude that
the present mean-field description predicts $\nu \simeq 1.5$ if
$P-P_c > 0.1$ and $ \nu \simeq 2.5 $ for the smaller value of $P-P_c$.

\begin{figure}
\centerline{\epsfxsize=8cm
            \epsfbox{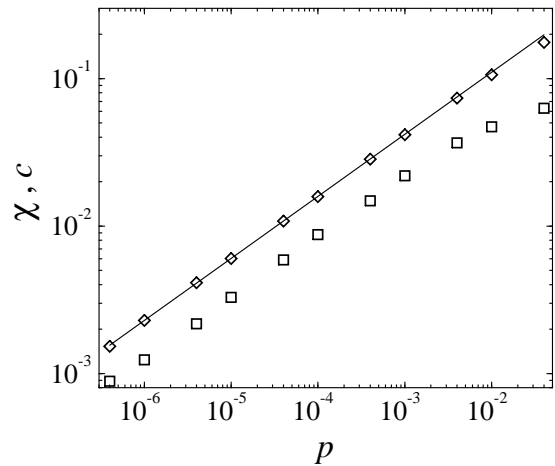}
            \vspace*{1mm}   }
\caption{MC results for the particle-antiparticle concentration (open
diamonds) and its fluctuation (open squares) as a function of $p$ in
the BARW model described in the text. The solid line (slope 0.42)
indicates a fitted power law.}
\label{fig:papv3}
\end{figure}

In order to check the role of fluctuations we have performed MC
simulations on a particle-antiparticle BARW model. The system
evolution is governed by nearest neighbor jumps, particle-antiparticle
pair annihilations and creations as follows. A randomly chosen particle
(or antiparticle) can create a particle-antiparticle pair located on two
randomly chosen nearest neighbor sites with a probability $P_{br}$,
otherwise this particle jumps to one of the nearest neighbor sites.
The processes which would result in two particles (or antiparticles)
residing on the same point are blocked. A particle-antiparticle pair
is annihilated if they would stay on the same point as a result of the
mentioned elementary processes.
The branching rate $P_{br}$ is determined for a given particle as a
product $P_{br}=p R_1 R_2 R_3$, where $R_1$ ($R_2$, $R_3$) denotes the
distance between the chosen particles and its first (second, third)
neighbor antiparticles. The choice of this branching rate is motivated by
the topological fact that, in the original voter model, a vortex is connected
directly to three antivortices by boundary lines (see Fig.~\ref{fig:map3c}).
During the simulations the value of $P_{br}$ can become larger than 1
very rarely thus we did not need to reduce the time unit in which each
particle has a chance to create offsprings or to jump.
Initially particles and antiparticles with equal numbers are distributed
randomly on a square lattice. The system size is varied from $L=100$
to 400 when decreasing $p$. During the simulations we have recorded the
number of particles and determined the average value of the concentration
$c$ and its fluctuation $\chi$ in the stationary state as defined above.
From these MC results (see Fig.~\ref{fig:papv3}) we could confirm that
the concentration follows a power law (solid line) with an exponent
$\beta = 0.42(3)$. It is emphasized that this value of $\beta$ agrees
very well with the prediction of the above mean-field formalism
($\beta_{MF}=0.4$ for $\nu = 1.5$). This result implies that any
value of the exponent $\beta$ is reproducible with the parameter
adjustment of a more sophisticated BARW model.

Considering the fluctuations obtained by simulations, a striking
difference is found between the present BARW model and the
three-candidate voter model. The BARW simulations (squares in
Fig.~\ref{fig:papv3}) indicate that $\chi (p) \propto c(p)$, which
seems to be a typical behavior for the BARW models \cite{LC,BARW2D}.
For the voter model, on the contrary, $\chi$ diverges
(see Fig.~\ref{fig:v3cf}) for small values of the control
paramater. We can identify two possible sources for this discrepancy.
First: during the diffusive motion the particle-antiparticle
annihilation results in an aggregation of the two species \cite{TW},
whereas in the vortex-antivortex system such a process is strongly
limited by the mentioned topological features. Second:
the three-fold degeneracy of the absorbing states of the three-candidate
voter model has no counterpart in the traditional BARW models.
Notice that, in the suggested BARW models, the evolution into an 
absorbing state is prevented by the divergency of the branching rate
in the limit $c \to 0$ for sufficiently large system size. Further
systematic research is required to clarify the effect of these phenomena.

To summarize, in the present paper we have improved the accuracy of
the numerical analysis of the critical transition appearing in the
vortex density for the three-candidate voter model when varying the
magnitude of cyclic dominance. Recognizing that the dynamics of the
vortices and antivortices is similar to a BARW model with a 
density-dependent particle-antiparticle pair creation (branching) rate, we
have contrasted these two systems. According to our comparison, we
can state that the power law behavior of the vortex density is
reproducible with a suitable choice of the pair creation mechanism. 
The same is not true for the behavior of fluctuations, which seems to be 
quite different in the two models. This discrepancy is a
 motivation to seek further extensions of BARW models, since this approach 
seems to be very useful in the investigation of the self-organizing,
 three-color domain structures.

\acknowledgements
We thank P. Krapivsky for a critical reading of the manuscript.
G. S. acknowledges a senior research fellowship from PRAXIS (Portugal).
Supports from NATO (CRG-970332), PRAXIS (project PRAXIS/2/2.1/Fis/299/94)
and the Hungarian National Research Fund (T-23552) are also acknowledged.

\end{document}